\documentclass{moriond}

\usepackage{amssymb,amsmath,bm}

\def\be{\begin{equation}}
\def\ee{\end{equation}}
\def\bea{\begin{eqnarray}}
\def\eea{\end{eqnarray}}

\newcommand{\nt}{\widetilde{\chi}^0}
\newcommand{\ch}{\widetilde{\chi}^\pm}
\newcommand{\chp}{\widetilde{\chi}^+}
\newcommand{\chm}{\widetilde{\chi}^-}
\newcommand{\stau}{\widetilde{\tau}}
\newcommand{\ord}[1]{\mathcal{O}\left( #1 \right)}

\newcommand{\Photo}{}

\begin{document}


\vspace*{4cm}
\title{A lower bound on light neutralino dark matter from LHC data}

\author{L.~Calibbi $^{\star}$~\footnote{Speaker},
J.~M.~Lindert $^{\dagger}$,
T.~Ota $^{\ddag}$,
Y.~Takanishi $^{*}$}
	
\address{
$^{\star}$~Service de Physique Th\'eorique, Universit\'e Libre de Bruxelles, B-1050 Brussels, Belgium\\
$^{\dagger}$~Physik-Institut,  Universit\"at Z\"urich, CH-8057 Z\"urich,  Switzerland \\
$^{\ddag}$~Department of Physics, Saitama University, 338-8570 Saitama-Sakura, Japan\\
$^*$~Max-Planck-Institut f\"ur Kernphysik, D-69117 Heidelberg, Germany
}

\maketitle

\abstracts{Under the hypothesis that the MSSM neutralino accounts for the observed dark matter density,
we investigate how light this particle is still allowed to be after the latest LHC data.
In particular, we discuss the impact of searches for events with multiple taus and missing transverse momentum, 
which are a generic prediction of the light neutralino scenario.}

\section{Introduction}

The LHC collaborations have performed a large number of searches for supersymmetric particles employing 
the first run data at $\sqrt{s}=7+8$~TeV.
A remarkable outcome of these analyses is that they started to test directly the electroweak sector of the theory,
setting in several cases constraints on the purely electroweakly interacting new particles -- such as sleptons, charginos and neutralinos -- 
way beyond the previous bounds set by LEP. This allows us to probe the parameter space of the Minimal Supersymmetric Standard Model (MSSM) 
that features a neutralino as a perfect dark matter (DM) candidate, since the properties of electroweakly interacting SUSY particles have 
a crucial role 
in the determination of the neutralino relic density. In particular, the question we want to answer to is: 
\begin{center}
{\it how light can neutralino dark matter be after the $\sqrt{s}=7+8$ TeV run data?} 
\end{center}
We address here this problem in a framework defined by the following assumptions: 
(i) the field content of the MSSM only; (ii) conserved R-parity and a neutralino as the lightest supersymmetric particle; 
(iii) a standard thermal history of the universe, that implies that the neutralino relic density has to 
fulfill the upper bound from CMB observations, which reads at 3$\sigma$:~\cite{Ade:2013zuv} 
\begin{equation}
\Omega_{\rm DM}h^2 \le 0.124 \,. \label{eq:planck}
\end{equation}
As we are going to discuss, the above requirement translates into certain conditions on the SUSY spectrum, in particular implying that some 
particles must be relatively light, hence giving a handle to test light neutralino dark matter at the LHC.

\section{Light neutralino dark matter in the MSSM}

Since we are interested in light dark matter (much below 100 GeV) and chargino searches at LEP constrain Wino and Higgsino 
masses to be $M_2,~\mu\gtrsim 100$ GeV,
the lightest neutralino $\nt_1$ must be mainly Bino. 
A Bino-like neutralino is typically overproduced in thermal processes in the early universe so that an efficient annihilation mechanism 
is needed in order to reproduce the observed relic abundance. As a consequence, the bound in (\ref{eq:planck}) selects
specific regions of the SUSY parameter space and thus sets certain conditions on the spectrum.
In principle, there are several ways to achieve a large enough annihilation cross-section, 
but if we concentrate on the following mass range for the neutralino:
\begin{equation}
m_{\nt_1}\lesssim 30~{\rm GeV},
\label{eq:lspmass}
\end{equation}
the number of options drastically reduces. 
\begin{figure}[t]
\centering
\includegraphics[width=0.8\textwidth]{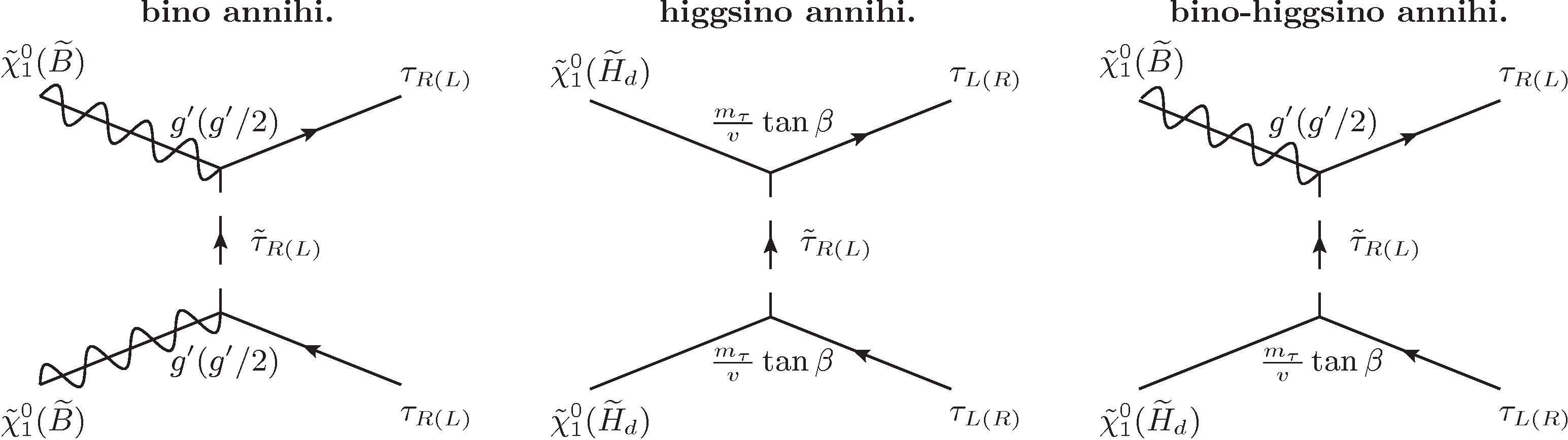}
\caption{\label{fig:ann}
Relevant neutralino annihilation processes mediated by a light stau.}
\end{figure}
It has been shown,~\cite{AlbornozVasquez:2011yq,Grothaus:2012js,Calibbi:2013poa} 
that the conditions in (\ref{eq:planck}, \ref{eq:lspmass}) can be satisfied if the annihilation of neutralinos
takes place through the t-channel exchange of a tau slepton. A combination of LEP, LHC and indirect observables exclude all
other possibilities, for instance the mediation of a 100$\div$200 GeV CP-odd scalar.~\cite{Calibbi:2011ug}
The only other possibility left~\cite{Arbey:2012na} is a scenario with a very light sbottom and an $\ord{1}$ GeV mass splitting 
between the neutralino and a sbottom -- in order to evade sbottom searches at LEP -- as well as a tuned left-right sbottom mixing such that the $Z$-$\widetilde{b}$-$\widetilde{b}$ interaction is strongly reduced, so that the $Z$ width constraints do not apply.
Besides this corner of the parameter space that is very difficult to test, the 
relic density bound on light neutralino DM implies the presence of a stau with a mass of few hundreds GeV at most. 
The diagrams in figure \ref{fig:ann} show that a right-handed stau is preferred, having twice the hypercharge of the left-handed one, and the
process gets more efficient if the Yukawa interaction Higgsino-tau-stau contributes. For that to occur, a sizeable Higgsino component
in $\nt_1$ is required and thus the Higgsinos cannot be too heavy. Furthermore, a large value of $\tan\beta$ is preferred.\footnote{
A further contribution, not shown in figure {\ref{fig:ann}}, is given by a mixed left-right stau exchange. In this case, 
the annihilation can occur in the s-wave without the need of a chirality flip in an external tau line. However,
it is not as effective as the other contributions discussed here in the case of very light neutralinos.~\cite{stauL}}

The consequence of the above discussion is twofold: (a) the constraint on the relic abundance implies the presence of further light particles,
the right-handed stau, two Higgsino-like neutralinos and charginos; (b) the framework is completely defined by a handful of parameters that
controls the calculation of the annihilation cross-section. We can thus work within a simplified setup described by 
\be
m_{\widetilde{\tau}_R},~M_1,~\mu,~\tan\beta,
\label{eq:parameters}
\ee
that are respectively the soft SUSY-breaking stau and Bino masses, the supersymmetric Higgsino mass and the usual ratio of vevs that 
controls the size of the Yukawas couplings of the leptons. All other SUSY parameters can be neglected, in particular the other
superpartners might be in principle much heavier than those involved in the DM annihilation. 
\begin{figure}[t]
\centering
\includegraphics[width=0.45\textwidth]{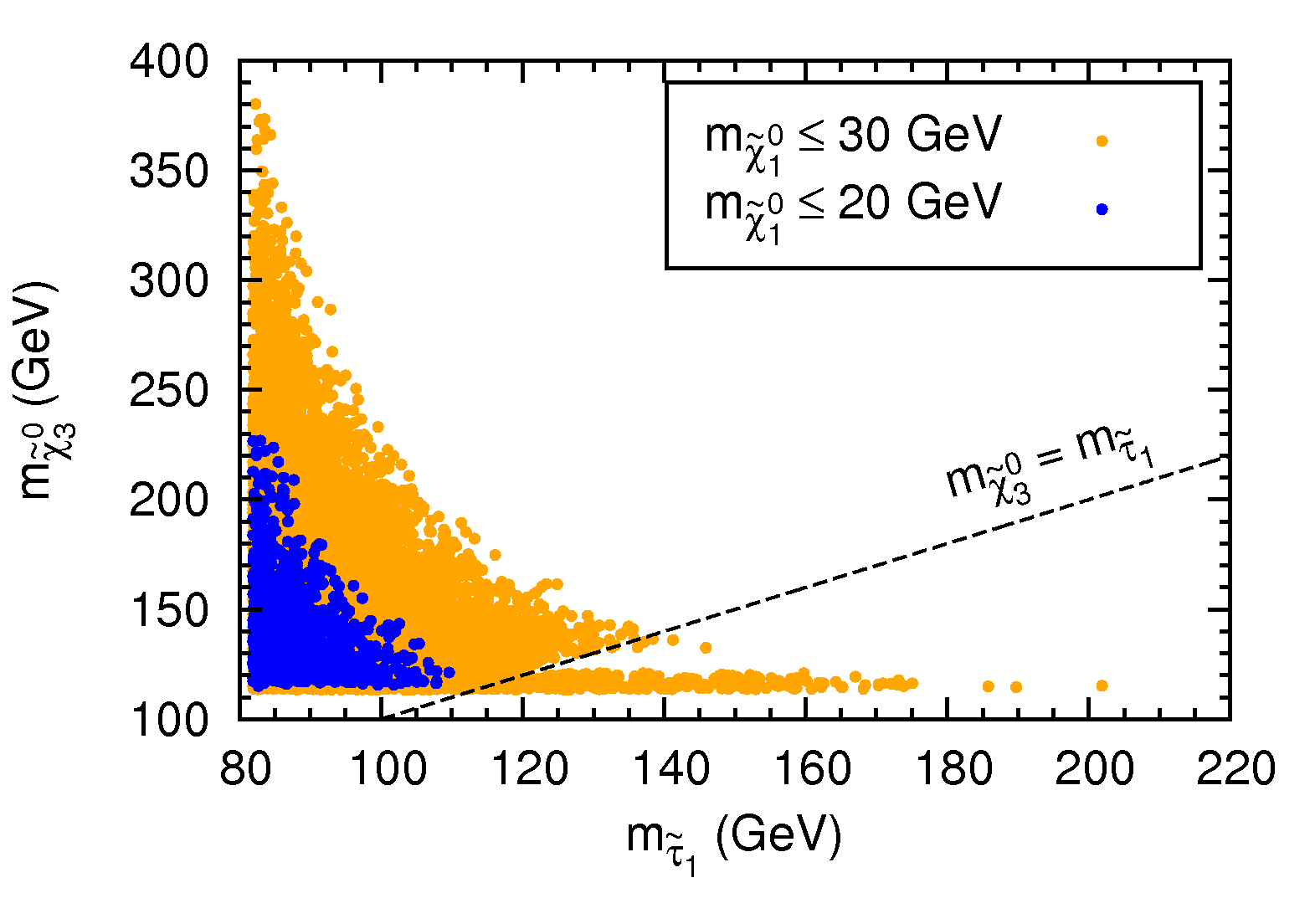}
\includegraphics[width=0.45\textwidth]{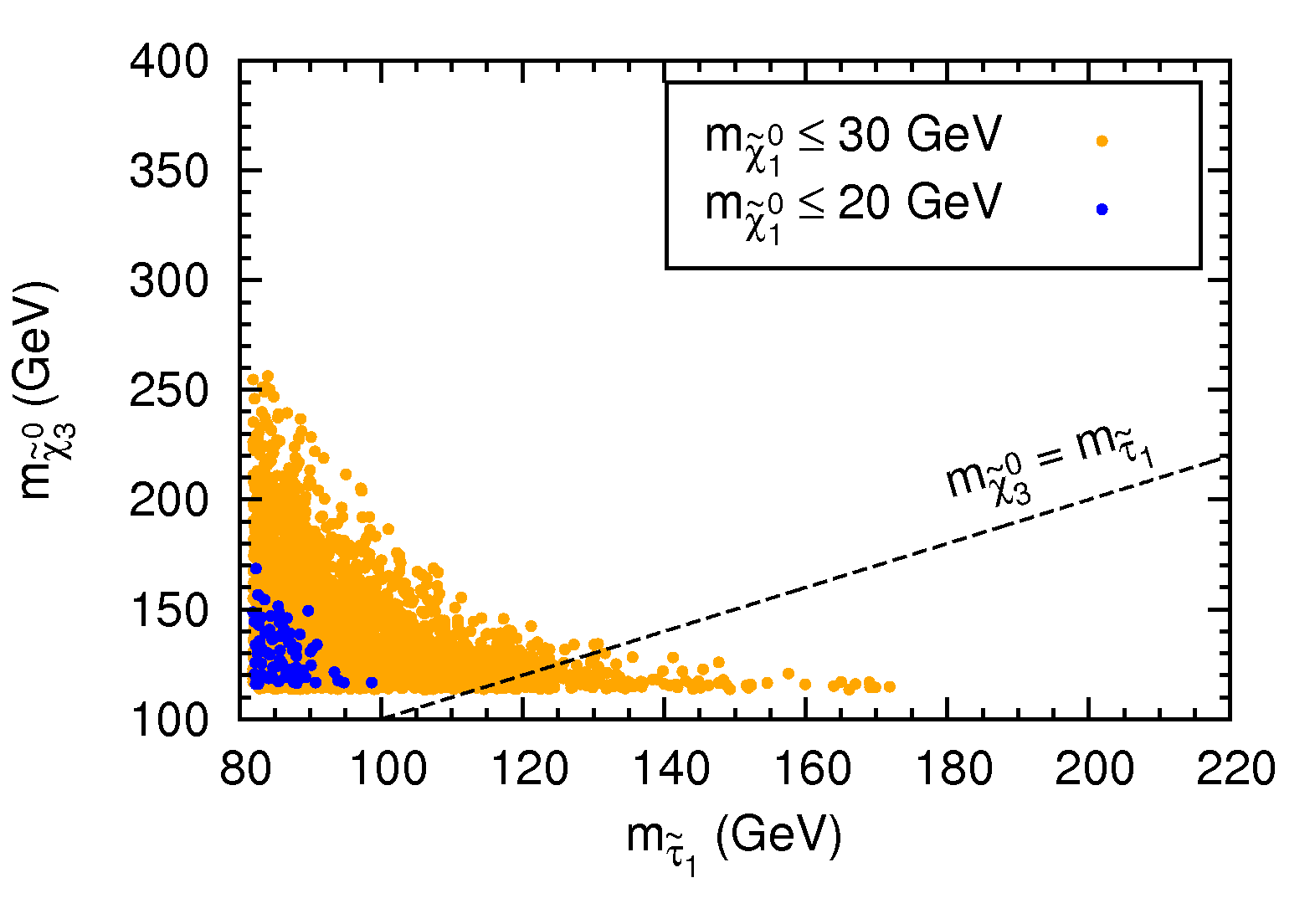}
\caption{Points with a light neutralino DM displayed on the $m_{\stau_1}$-$m_{\nt_3}$
  plane for $\mu>0$ (left) and $\mu<0$ (right).\label{fig:parspace}}
\end{figure}

We performed a scan of the above defined parameter space,~\cite{Calibbi:2013poa} applying all the relevant phenomenological constraints,
including LEP measurements of the invisible $Z$ width and limits from direct searches for staus, charginos and neutralinos
at LEP.
The result is displayed in figure \ref{fig:parspace}, where the points that satisfy (\ref{eq:planck}, \ref{eq:lspmass})
are shown in the plane of the physical masses of the lightest stau and the heaviest Higgsino-like neutralino.\footnote{We remind
that, being all controlled by $\mu$, $m_{\ch_1}\approx m_{\nt_2} \approx m_{\nt_3}$.} 
The points in blue indicate $m_{\nt_1}\le 20$ GeV, in particular they can feature a neutralino as light as $10\div 12$ GeV.
The left panel corresponds to $\mu>0$ (left), the right one to  $\mu<0$ (right).
From the figure we see that, as anticipated, both $\stau_1$ and the Higgsinos have to be rather light in order
the relic density constraint to be fulfilled by a light neutralino. In particular:
\begin{equation}
  m_{\stau_1}\lesssim 200~{\rm GeV}, \quad m_{\nt_3}  \lesssim 400 ~{\rm GeV}.
\label{eq:up-bounds}
\end{equation}
Comparing the two panels, we can also see that satisfying the bound of (\ref{eq:planck}) 
for $\mu<0$ (right panel) requires somewhat lighter Higgsinos. 
This is due to a destructive interference among the diagrams in figure \ref{fig:ann}.

Before moving to the next section and discussing how this spectrum can be tested at the LHC, let us briefly mention further possible 
constraints from DM experiments (direct and indirect). 
The prediction for the spin-independent scattering cross-section with nuclei is affected by the uncertainties of light quark masses
and hadronic matrix elements. Moreover, it substantially depends on parameters that do not enter the calculation of the relic abundance,
such as the CP-odd scalar mass and the squark masses. Taking this into account, the scattering cross-section in our
scenario approximately ranges between $10^{-44}$ and $10^{-46}$ cm$^2$.~\cite{Calibbi:2013poa} The most stringent limit, provided by LUX, 
is about $10^{-45}$ cm$^2$ in the DM mass range we are considering, $10~{\rm GeV}\lesssim m_{\nt_1} \lesssim 30~{\rm GeV}$.~\cite{LUX}
Therefore, we can conclude that the sensitivity of direct detection experiments started to reach the typical cross-section of a light neutralino DM 
without being capable of fully probing this scenario.
For what concerns indirect detection experiments, similar conclusions can be drawn. The most sensitive search for DM annihilation into taus
has been performed by the Fermi-LAT collaboration employing gamma-ray data from dwarf satellite galaxies. 
The limit they obtain on the annihilation cross-section is $\langle \sigma_{\rm ann.} v\rangle \gtrsim 2\times 10^{-26}$ cm$^3$/s,~\cite{fermi} 
while the prediction of our scenario typically ranges between  $10^{-26}$ and $10^{-27}$ cm$^3$/s.

\section{LHC tests of light neutralinos}

As discussed above, the relic density constraint in (\ref{eq:planck}) translates to a well-defined spectrum, where at least the
following states must be lighter than few hundreds GeV:
\begin{equation}
\stau_1,~\nt_2,~\nt_3,~\ch_1. 
\label{eq:spectrum}
\end{equation}
The rest of the spectrum is not constrained by the DM relic abundance and can be in principle heavier. 
As a consequence a model-independent test of the light neutralino parameter space
must rely on direct electroweak production of the above-listed particles.  
This takes place through the electroweak Drell-Yan mechanism (s-channel $Z/\gamma$ exchange) and the relevant modes are:
\begin{align}
  pp \to \stau^+_1\, \stau^-_1 +X,\quad pp \to \nt_i\, \nt_j + X ,\quad
  pp \to \nt_i\, \ch_1 +X,\quad pp \to \chp_1\, \chm_1 +X ,
\end{align}
where $i,j=2,3$.\footnote{In principle, production of $\nt_1\nt_1$ could be also searched for in mono-jet + missing $E_T$ events. 
The sensitivity of this channel at $\sqrt{s}=8$ TeV is however too low to set any constraint.~\cite{mono}} 
The subsequent decays depend on the hierarchy of the particles listed in (\ref{eq:spectrum}).
If $m_{\ch_1} \simeq m_{\nt_{2,3}} > m_{\stau_1} > m_{\nt_1}$, as it occurs in most of the plane shown in figure \ref{fig:parspace},
the Higgsino-like neutralinos dominantly decay to a tau and an on-shell stau. The stau always decays to a tau and a neutralino. 
As a consequence, the resulting decay chain is:
\begin{align}
 \nt_{2,3}  &\to  \tau^\pm \stau_1^\mp  \to \tau^+ \tau^- \nt_1.
\end{align}
As a consequence, production of Higgsino pairs, e.g.~$\nt_2\nt_3$ and $\nt_{2,3}\ch_1$, gives final states with 4 or 3 taus 
plus missing $E_T$. Multi-tau events are therefore a generic feature of light neutralino DM and thus a very powerful handle to test
this scenario. In section \ref{sec:LHC}, we are going to show how a recent ATLAS search for events with multiple 
hadronically decaying taus can be employed to set a strong constrain on the light neutralino parameter space.
On the other hand, in the corner of the plane of figure \ref{fig:parspace} where the Higgsinos are lighter than the stau, 
we expect a reduced sensitivity. 
This case is however strongly constrained by the limits posed by LHC experiments to invisible decays of the scalar, 
to which $h\to\nt_1\nt_1$ contributes in our case, as we discuss in the following. 

\subsection{Invisible h decay}
\begin{figure}[t]
\centering
 \includegraphics[width=0.45\textwidth]{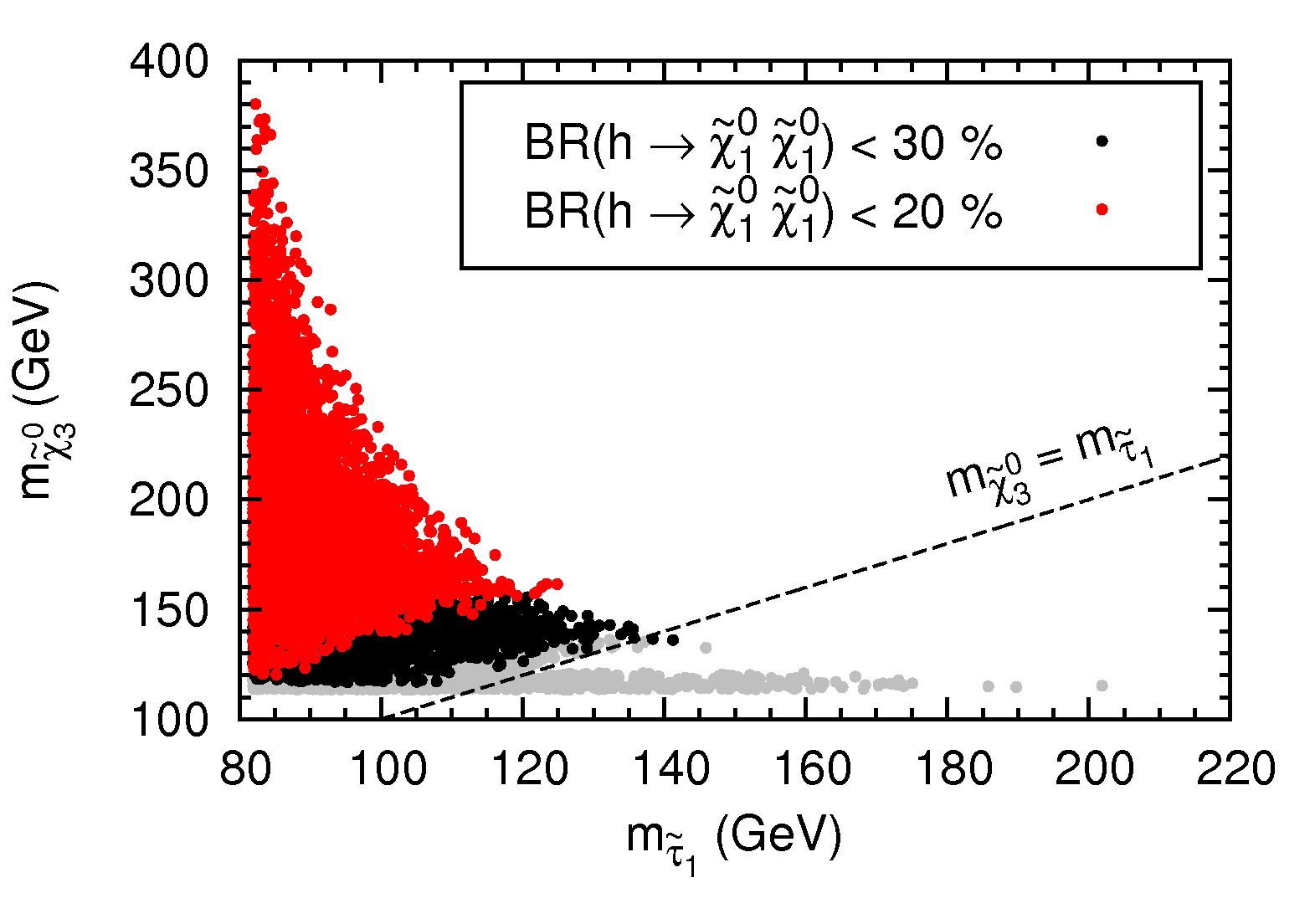}
 \includegraphics[width=0.45\textwidth]{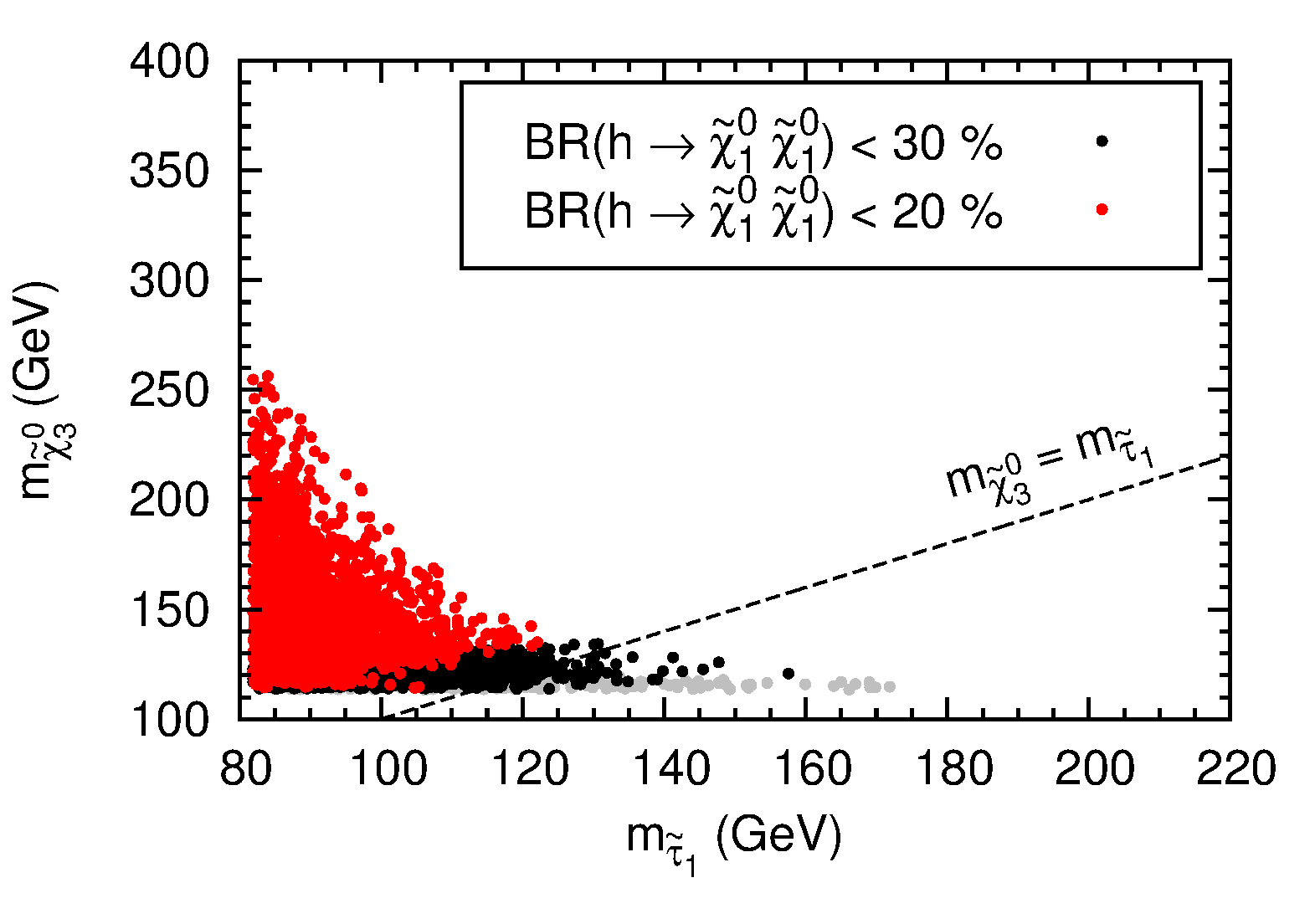}
 \caption{Different values for BR($h\to \nt_1\nt_1$) on the
  $m_{\stau_1}$-$m_{\nt_3}$ plane for $\mu>0$ (left) and $\mu<0$ (right).
\label{fig:hchichi} }
\end{figure}

As discussed above, the relic density bounds requires
sizeable Higgsino components in $\nt_1$. Thus the neutralino
can couple with $h$ and large branching ratios for the decay $h\to
\nt_1\nt_1$ can be induced. The lighter the Higgsinos, the larger
we expect BR($h\to \nt_1\nt_1$) to be. The decay width is given by
\begin{align}
\label{eq:hchichi}
& \Gamma (h \rightarrow \nt_1 \nt_1) = \frac{G_F M_W^2 m_h}{2 \sqrt{2} \pi}~
\left(  1- \frac{4 m_{\nt_1 }^2}{m_h^2}   \right)^{3/2} \big\vert  C_{h \nt_1 \nt_1 }   \big\vert^2 \;,\\
 & C_{h \nt_1 \nt_1 } = \big( N_{12} -\tan \theta_W \; N_{11}   \big)
\big( \sin\beta\; N_{14} -\cos\beta \; N_{13}   \big)\;,
\end{align}
where $N_{1i}$ refer to the gaugino/Higgsino components in the neutralino:
$\nt_1 = N_{11} \widetilde{B} + N_{12} \widetilde{W} + N_{13} \widetilde{H}_d + N_{14} \widetilde{H}_u$.
In the limit $M_2\gg \mu$, the Higgsino components can be simply written as:
\begin{equation}
N_{13} = 
\frac{M_{Z} s_{W}^{}}{\mu}
\left[
 s_{\beta} + c_{\beta} \frac{M_{1}}{\mu}
 \right],
 \quad
 N_{14} =
 -
 \frac{M_{Z} s_{W}^{}}{\mu}
 \left[
 c_{\beta}
 +
 s_{\beta}
 \frac{M_{1}}{\mu}
 \right]. 
 \label{eq:higgsino}
\end{equation}
As we can see, these parameters -- and thus $\Gamma(h \rightarrow \nt_1 \nt_1)$ -- grow for a smaller Higgsino mass $\mu$, as expected. 
Moreover, there is a non-trivial dependence on the relative sign of $M_1$ and $\mu$. Choosing $M_i>0$ without loss of generality,
we then see that for $\mu>0$ we expect a larger effect than for $\mu<0$, for which a partial compensation among the terms in 
(\ref{eq:higgsino}) occurs.
The resulting prediction for  BR($h\to \nt_1\nt_1$) is shown in figure \ref{fig:hchichi}. 
We observe a reduction of the invisible width in the $\mu<0$ case, as expected from the above mentioned cancellations in (\ref{eq:higgsino}).
However, the branching ratio is sizeable for light Higgsinos in both cases.
As we can see, one
always finds BR($h\to \nt_1\nt_1$)$>$30\,(20)$\%$ for $\mu>0$ ($\mu<0$) 
in the region where multi-tau searches are kinematically disfavoured because $\nt_{2,3}  \to  \tau \stau_1$ can not occur on shell. 
Hence this region is strongly disfavoured by the fits of 
${\rm BR}_h^{\rm inv}\equiv{\rm BR}(h \to {\rm invisible})$ 
to the observed decay rates, that give:~\cite{Giardino:2013bma}
\begin{align}
\label{eq:hinv}
 {\rm BR}_h^{\rm inv} \lesssim 20 \%\quad(95\%~{\rm CL}). 
\end{align}

\subsection{Multi-tau limits}
\label{sec:LHC}
\begin{figure}[t]
\centering
 \includegraphics[width=0.45\textwidth]{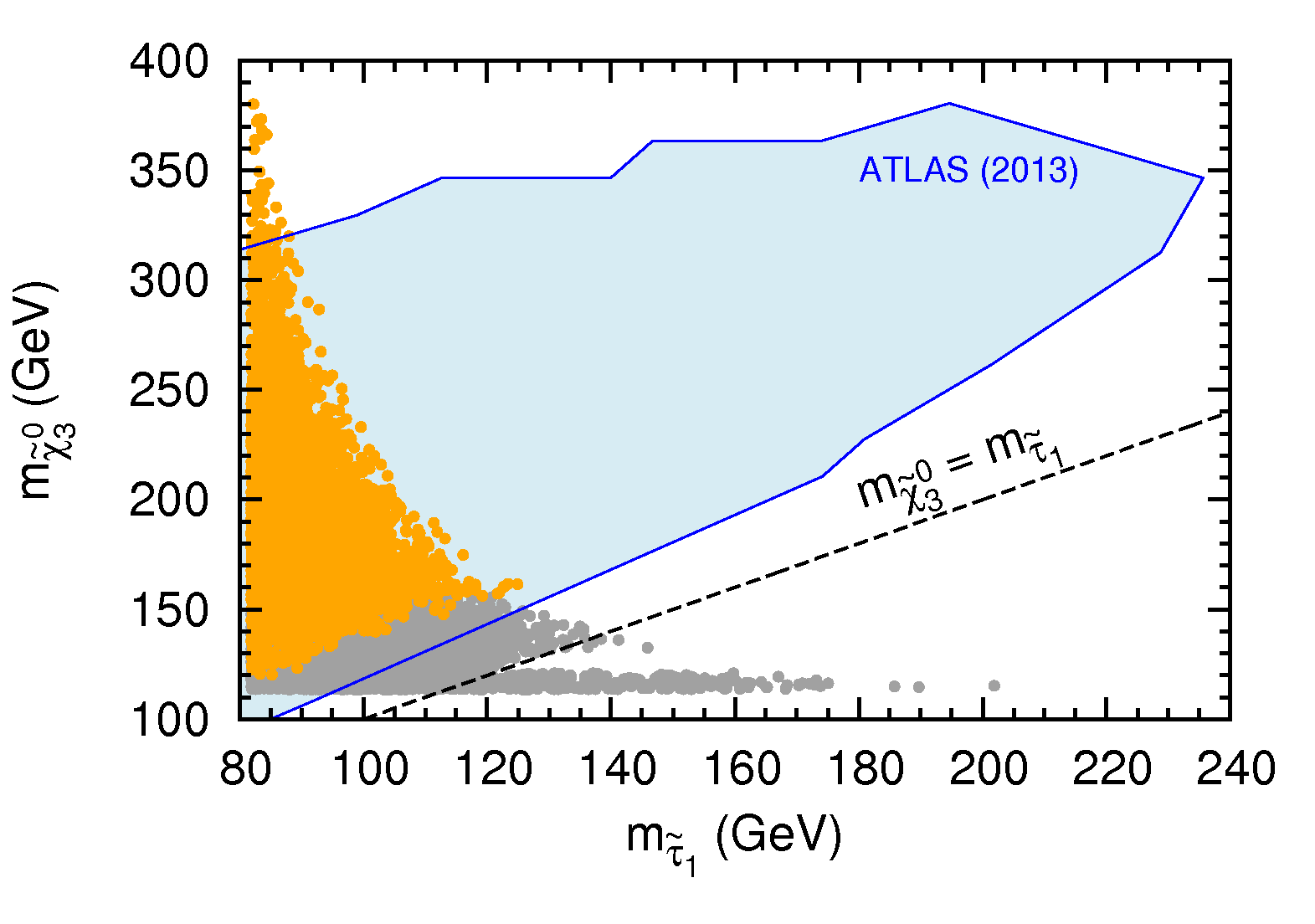}
 \includegraphics[width=0.45\textwidth]{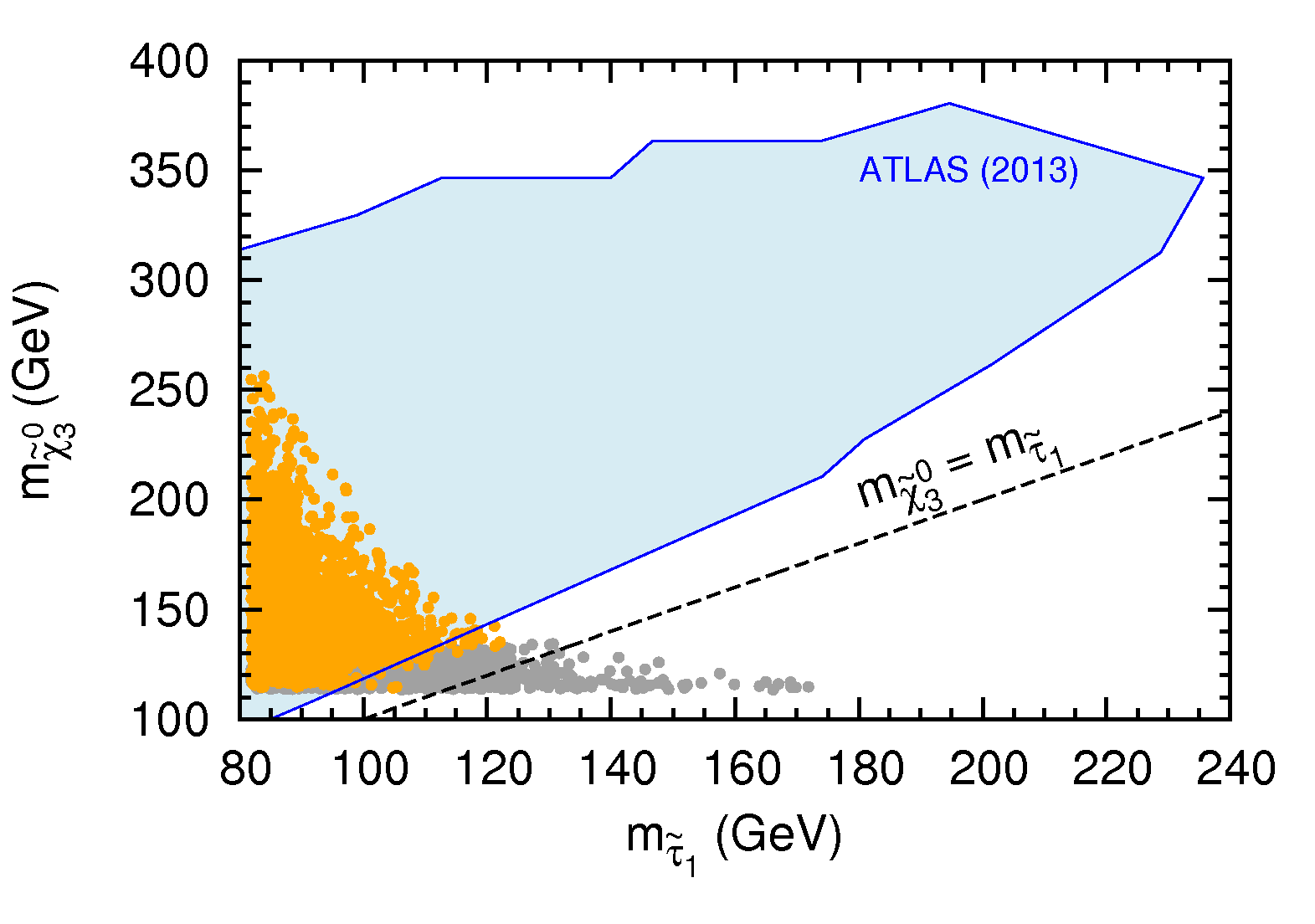}
 \caption{ATLAS multi-tau limit on our 
  $m_{\stau_1}$-$m_{\nt_3}$ plane for $\mu>0$ (left) and $\mu<0$ (right).
   Orange points correspond in addition to BR($h\to \nt_1\nt_1$)$<$20\%.
 \label{fig:atlas}}
\end{figure}

As we have discussed above, the spectrum predicted by the light neutralino DM scenario can be tested at the LHC through 
multi-tau plus missing $E_T$ events. 
The ATLAS collaboration has recently performed a
search for new physics in a final state with at least two hadronically decaying taus and
large missing transverse momentum
employing $20~{\rm fb}^{-1}$ of data at $\sqrt{s}=8$~TeV.~\cite{ATLAS:2013yla} 
The null result has been interpreted by the collaboration in different simplified models. We have recast the limits
into our parameter space,~\cite{Calibbi:2013poa} as defined in (\ref{eq:parameters}). 
In order to do so, we generated the signal events employing {\tt Herwig ++},~\cite{Bahr:2008pv} computed the K-factors with {\tt Prospino 2}~\cite{Beenakker:1999xh} 
and simulated the detector response by means of {\tt Delphes 3}.~\cite{Ovyn:2009tx} We applied the same cuts and defined the same signal
regions as the ATLAS analysis.~\cite{ATLAS:2013yla} We validated our analysis by reproducing the ATLAS exclusion on the  $M_2$-$\mu$
plane.~\cite{Calibbi:2013poa} We could then estimate the limit in our parameter space. The result is shown in figure \ref{fig:atlas}.
As we can see, the ATLAS search sets a very strong constraint on the portion of the plane where the Higgsinos can decay to a real stau,
excluding Higgsinos up to $\approx 380$ GeV and staus up to $\approx 230$ GeV.
Very few points consistent with neutralino DM escape this bound, especially if we combine it 
with the limit on the invisible $h$ branching ratio (\ref{eq:hinv}):
for $\mu>0$, those lying in the corner with heavy Higgsinos 
$\gtrsim 320$ GeV and a stau mass very close to the LEP bound; for $\mu<0$, some points that feature a small Higgsino-stau
mass splitting -- thus giving soft taus in the final state -- with ${\rm BR}_h^{\rm inv} \approx 20$ \%.
In terms of the neutralino mass, the limit we obtained is:
\begin{equation}
 m_{\nt_1}\gtrsim 24~(22)~{\rm GeV}.
\end{equation}
for the case $\mu>0$ ($\mu<0$). 
Clearly, a mild improvement of the sensitivity of multi-tau searches and of the upper bound on ${\rm BR}_h^{\rm inv}$ 
would probe the surviving corners of figure \ref{fig:atlas}, thus fully testing light neutralino DM up to 30 GeV.

\section{Conclusions}

We discussed how searches for electroweakly interacting SUSY particles can be employed to constrain
models with light neutralino DM.  
We have shown that neutralinos with a mass up to 30 GeV are good DM candidates only provided that
staus and Higgsinos have masses of the order of few hundreds GeV. This is a consequence of the requirements
set to the spectrum by the observed relic abundance. As the calculation of the relic density depends on few parameters,
we could define a manageable simplified model to study the light neutralino parameter space. 
The obtained spectra generically predict cascade decays resulting in multi-tau signals. 
The reinterpretation of an ATLAS search, in combination with information on $h\to$ invisible, allowed us
to set a constraint on the neutralino DM parameter space that is tighter than those from direct and
indirect DM searches. The results for the case $\mu<0$, that was not considered before because disfavoured by 
the anomalous magnetic moment of the muon, were presented here for the first time. The outcome is 
qualitatively similar to the case with positive $\mu$ that we previously studied.

\section*{Acknowledgments}

L.C.~would like to thank the Moriond organizers for the financial support and the opportunity of presenting this talk.

\end{document}